\documentclass[journal]{IEEEtran}

\usepackage{cite}
\usepackage{amsmath,amssymb,amsfonts}
\usepackage{algorithmic}
\usepackage{algorithm}
\usepackage{graphicx}
\usepackage{textcomp}
\usepackage{xcolor}
\usepackage{booktabs}
\usepackage{multirow}
\usepackage{array}
\usepackage{url}
\usepackage{pifont}
\usepackage{hyperref}

\begin{document}

\title{AMDS: Attack-Aware Multi-Stage Defense System for Network Intrusion Detection with Two-Stage Adaptive Weight Learning}

\author{
\begin{tabular}{cc}
\textbf{Oluseyi Olukola} & \textbf{Nick Rahimi} \\
\parbox[t]{3in}{\centering School of Computing Sciences and \\ Computer Engineering} & 
\parbox[t]{3in}{\centering School of Computing Sciences and \\ Computer Engineering} \\
University of Southern Mississippi & University of Southern Mississippi \\
Hattiesburg, MS, USA & Hattiesburg, MS, USA \\
\texttt{oluseyi.olukola@usm.edu} & \texttt{nick.rahimi@usm.edu}
\end{tabular}
}

\maketitle

\begin{abstract}
Machine learning based network intrusion detection systems are vulnerable to adversarial attacks that degrade classification performance under both gradient-based and distribution shift threat models. Existing defenses typically apply uniform detection strategies, which may not account for heterogeneous attack characteristics. This paper proposes an attack-aware multi-stage defense framework that learns attack-specific detection strategies through a weighted combination of ensemble disagreement, predictive uncertainty, and distributional anomaly signals. Empirical analysis across seven adversarial attack types reveals distinct detection signatures, enabling a two-stage adaptive detection mechanism. Experimental evaluation on a benchmark intrusion detection dataset indicates that the proposed system attains 94.2\% area under the receiver operating characteristic curve and improves classification accuracy by 4.5 percentage points and F1-score by 9.0 points over adversarially trained ensembles. Under adaptive white-box attacks with full architectural knowledge, the system appears to maintain 94.4\% accuracy with a 4.2\% attack success rate, though this evaluation is limited to two adaptive variants and does not constitute a formal robustness guarantee. Cross-dataset validation further suggests that defense effectiveness depends on baseline classifier competence and may vary with feature dimensionality. These results suggest that attack-specific optimization combined with multi-signal integration can provide a practical approach to improving adversarial robustness in machine learning-based intrusion detection systems.
\end{abstract}

\begin{IEEEkeywords}
Network intrusion detection, adversarial machine learning, ensemble learning, adaptive defense, attack-specific optimization
\end{IEEEkeywords}


\section{Introduction}
\label{sec:introduction}

Network Intrusion Detection Systems (NIDS) have become a cornerstone of modern cybersecurity infrastructures, enabling real-time detection of malicious network traffic in increasingly complex threat landscapes. The adoption of machine learning (ML) and deep learning (DL) techniques has markedly improved the ability of NIDS to detect both known and zero-day attacks~\cite{wang2024zero} by learning intricate traffic patterns beyond what static signature-based systems capture~\cite{buczak2016survey,vinayakumar2019deep,ahmad2021network}. This progress spans diverse architectural paradigms, including graph neural networks for network flow analysis~\cite{chen2023graph}, transformer-based sequence modeling~\cite{liu2024transformer}, deep learning approaches for volumetric attack detection~\cite{ferrag2023deep}, and federated learning for privacy-preserving distributed deployment~\cite{nguyen2023federated}. Despite these advances, ML-based NIDS remain vulnerable to adversarial attacks carefully crafted perturbations to network traffic that are imperceptible to human analysts yet capable of substantially degrading detection performance~\cite{madry2017towards,venturi2024,ennaji2024,sharma2024}.

Recent studies have identified multiple categories of adversarial attacks against NIDS~\cite{ring2019survey,khraisat2019survey}. Gradient-based methods such as FGSM and PGD perturb input features to evade detection while maintaining minimal distortion~\cite{goodfellow2014explaining,madry2017towards,szegedy2014intriguing}. Decision-based black-box attacks exploit model output spaces without requiring gradient information~\cite{carlini2017towards,papernot2016transferability,moosavi2016deepfool}. Distribution shift attacks including feature injection and traffic morphing alter statistical properties of network flows to bypass detection thresholds~\cite{zhao2024,venturi2024,lin2022idsgan}. Empirical studies suggest that gradient-based attacks can achieve 80--95\% attack success rates against NIDS models, while distribution shift attacks may evade detection in up to 72\% of cases~\cite{sharma2024,heydari2025,apruzzese2021hardening}.

Several defense paradigms have been proposed in response to these threats~\cite{sun2023adversarial}. \textit{Adversarial training} incorporates adversarial examples during model training, improving robustness against known attacks but often sacrificing clean-data accuracy and failing to generalize to unseen attack types~\cite{tramer2018ensemble,madry2017towards,biggio2013evasion}. \textit{Ensemble learning} combines multiple classifiers to leverage complementary model strengths~\cite{dietterich2000ensemble,barik2025,awad2025,liu2020towards}; however, existing ensemble approaches treat model disagreement as noise to be averaged away through voting schemes, potentially missing the signal that disagreement may indicate adversarial manipulation. \textit{Preprocessing and detection gates}, including diffusion-based input purification and uncertainty-aware anomaly detection, show promise in identifying adversarial examples~\cite{merzouk2024,xu2017feature,pang2020rethinking}, yet these methods typically rely on single detection signals that can be circumvented by adaptive adversaries~\cite{athalye2018obfuscated}.

Despite progress, existing defenses appear to face four recurring limitations~\cite{sharma2024,ennaji2024}: ensemble methods treat disagreement as noise rather than signal~\cite{dietterich2000ensemble}; current approaches tend to apply uniform detection strategies across heterogeneous attack types~\cite{xu2017feature,ma2018characterizing}; attack-specific defenses assume prior knowledge of attack type~\cite{dynamicids2024}; and systems typically provide either detection or classification, but not both. To address these limitations, we propose \textbf{AMDS (Attack-Aware Multi-Stage Defense System)}, integrating disagreement-triggered adaptive defenses, attack-specific weight learning, two-stage detection refinement, attack-adaptive model weighting, and dual-output architecture into a unified defense pipeline.

Specifically, this paper makes five contributions to adversarial defense for network intrusion detection.

\begin{enumerate}
\item We show that no single detection signal disagreement, entropy, or anomaly appears to provide adequate discriminative power for adversarial detection (individual AUCs: 0.138--0.549), and that combining three complementary signals through learned weighting yields 97.1\% AUC. This finding suggests that multi-signal integration with optimized weights may be a necessary foundation for effective adversarial defense.

\item We learn specialized detection weights optimized for each attack category through empirical analysis on six attack types (FGSM, PGD-L$_\infty$, C\&W, SPSA, Injection, Morphing), with PGD-L$_2$ additionally evaluated as a weak-effectiveness case (16.6\% attack success rate) to examine detector behavior under minimal adversarial signal. Our analysis indicates that gradient-based attacks tend to be detected through disagreement-dominant strategies (average $\beta=0.452$), while distribution shift attacks exhibit heterogeneous patterns (Morphing: anomaly-dominant $\gamma=0.533$, AUC=0.988; Injection: entropy-dominant $\alpha=0.433$, AUC=0.896). Building on these patterns, we propose two-stage adaptive detection combining generic inference with category-level refinement, which attains 94.2\% AUC (+36.3 points over generic-only, +17.2 over attack-specific-only approaches). Category inference achieves 100\% accuracy on evaluated attacks (7/7 correctly categorized), though this should not be extrapolated to broader attack taxonomies without further validation.

\item We introduce attack-adaptive model weighting where ensemble models are dynamically weighted based on inferred attack category. On evaluated attacks, this approach yields +30.0\,pp improvement on Injection and +17.9\,pp average on distribution attacks, with +11.0\,pp F1-score improvement overall. These gains appear particularly significant for distribution shift attacks, where baseline defenses including adversarial training provide minimal benefit.

\item We provide threat intelligence through a dual-output architecture delivering both adversarial detection (binary) and attack classification (multiclass), attaining 96.3\% classification accuracy on clean data and 84.8\% overall. To our knowledge, existing systems provide only one dimension of threat intelligence; AMDS is designed to deliver both simultaneously.

\item We conduct systematic evaluation against adversarially trained baselines (PGD training with three random seeds), finding that AMDS outperforms adversarial training by +4.5\,pp accuracy and +9.0\,pp F1-score, while adversarial training itself appears to provide minimal benefit over standard ensembles ($-0.6$\,pp accuracy). Under white-box adaptive attacks with full knowledge of the AMDS architecture, the system maintains 94.4\% accuracy with 4.2\% attack success rate, though this evaluation covers two adaptive variants and does not constitute a formal robustness guarantee. Cross-dataset evaluation on UNSW-NB15 suggests a deployment consideration: AMDS effectiveness appears to depend on baseline model competence, with attack strength scaling as $\sqrt{d}$ where $d$ is feature dimensionality.
\end{enumerate}

The remainder of this paper is organized as follows: Section~\ref{sec:related} reviews related work. Section~\ref{sec:method} presents the AMDS architecture. Section~\ref{sec:experiments} describes experimental methodology. Section~\ref{sec:results} reports results including component validation, baseline comparisons, cross-dataset generalization, and robustness analysis. Section~\ref{sec:conclusion} concludes with limitations and future directions.


\section{Related Work}
\label{sec:related}

Adversarial attacks against machine learning based network intrusion detection systems (NIDS) include gradient-based perturbations such as FGSM, PGD, and C\&W~\cite{goodfellow2014explaining,madry2017towards,carlini2017towards}, transfer-based black-box attacks~\cite{papernot2016transferability}, and problem-space manipulations that preserve protocol validity while altering feature distributions~\cite{venturi2024,usama2019generative,pawlicki2020towards}. Distribution shift attacks, including injection and morphing strategies~\cite{zhao2024,lin2022idsgan}, remain particularly challenging because robustness gained through adversarial training on gradient-based examples may not generalize effectively to these threats~\cite{zhong2020adversarial,deng2021adversarial}. Empirical studies report high attack success rates against NIDS models across both white-box and black-box settings~\cite{heydari2025,sharma2024}.

Defensive strategies can be broadly categorized into ensemble-based robustness and detection-based approaches. Ensemble adversarial training aims to reduce transferability across models~\cite{tramer2018ensemble}, while TRADES formulates a clean accuracy robustness trade-off~\cite{zhang2019trades}. In the NIDS domain, ensemble architectures incorporating deep learning~\cite{barik2025} and attention mechanisms~\cite{awad2025} improve resilience to gradient-based attacks, yet typically aggregate predictions through voting or averaging, implicitly suppressing disagreement rather than leveraging it as adversarial signal.

Detection-based defenses seek to identify adversarial inputs using individual signals such as feature inconsistencies~\cite{xu2017feature}, reconstruction error~\cite{meng2017magnet}, intrinsic dimensionality~\cite{ma2018characterizing}, uncertainty estimation~\cite{heydari2025}, or purification mechanisms~\cite{merzouk2024}. While effective under specific threat models, these approaches tend to rely on fixed thresholds and may be vulnerable to adaptive adversaries~\cite{athalye2018obfuscated}. Dynamic defense selection frameworks~\cite{dynamicids2024} introduce adaptive routing but depend on heuristic attack inference rather than learned optimization.

Despite significant progress, three recurring limitations can be identified in the existing literature. First, ensemble disagreement is typically treated as noise rather than as informative adversarial signal. Second, most defenses apply uniform strategies across heterogeneous attack types, which may lead to weaker robustness against distribution shift threats compared to gradient-based attacks. Third, existing systems generally provide either adversarial detection or attack classification, but not both simultaneously. The proposed framework seeks to address these limitations through learned attack-specific weight optimization, adaptive multi-stage detection spanning heterogeneous threats, and a dual-output architecture for integrated threat intelligence.


\section{Method}
\label{sec:method}

\begin{figure*}[!t]
\centering
\includegraphics[width=0.95\textwidth]{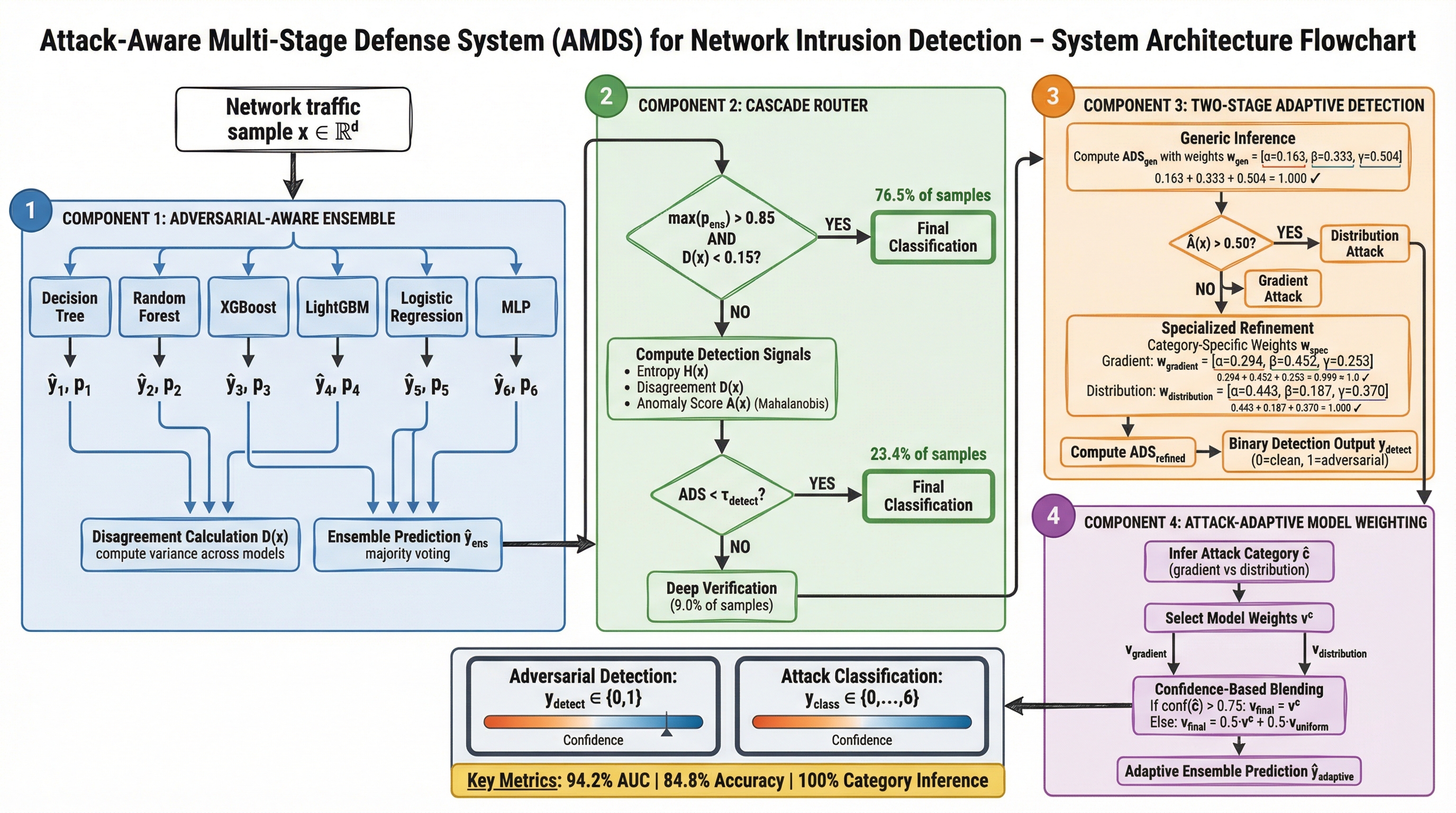}
\caption{AMDS system architecture. Network traffic is processed through four components: ensemble prediction with disagreement computation, cascade routing for efficiency-aware triage, two-stage adaptive detection with category-specific refinement, and attack-adaptive model weighting, producing dual outputs for adversarial detection and attack classification.}
\label{fig:system_architecture}
\end{figure*}

Figure~\ref{fig:system_architecture} illustrates the AMDS architecture, which integrates four components: an adversarial-aware ensemble (Section~\ref{sec:method_ensemble}), a cascade router for efficiency-aware triage (Section~\ref{sec:method_cascade}), a two-stage adaptive detector (Section~\ref{sec:method_ads}), and attack-adaptive model weighting (Section~\ref{sec:method_adaptive}). The following subsections detail each component, followed by the training and inference algorithms.

\subsection*{Adversarial-Aware Ensemble}
\label{sec:method_ensemble}

The adversarial-aware ensemble is designed to provide both base predictions and adversarial signals through model diversity.

\subsection{Ensemble Methodology}

\subsubsection{Model Selection and Disagreement}

We construct an ensemble $\mathcal{M} = \{M_1, M_2, \ldots, M_6\}$ of six classifiers spanning tree-based, boosting, linear, and neural paradigms to maximize architectural diversity (detailed configurations in Section~\ref{sec:experiments}). Each model $M_i$ produces a class prediction $\hat{y}_i \in \{0, 1, \ldots, C-1\}$ where $C=7$ classes, and a probability distribution $\mathbf{p}_i = [p_i^{(0)}, p_i^{(1)}, \ldots, p_i^{(C-1)}]$.

The ensemble prediction combines individual outputs through uniform voting:
\begin{equation}
\hat{y}_{\text{ens}} = \arg\max_{c} \sum_{i=1}^{6} \mathbb{1}[\hat{y}_i = c]
\label{eq:ensemble_prediction}
\end{equation}

Model disagreement serves as an adversarial signal, computed as the average per-class prediction variance:
\begin{equation}
D(\mathbf{x}) = \frac{1}{C} \sum_{c=0}^{C-1} \text{Var}(\{p_1^{(c)}, p_2^{(c)}, \ldots, p_6^{(c)}\})
\label{eq:disagreement}
\end{equation}

The intuition is that adversarial perturbations may affect diverse models unevenly, producing elevated disagreement relative to clean inputs.


\subsection*{Cascade Router}
\label{sec:method_cascade}

The cascade router implements a three-stage processing pipeline designed to route samples through increasingly sophisticated detection mechanisms only as needed.

In Stage~1 (fast ensemble screening), samples with high prediction confidence ($\max(\mathbf{p}_{\text{ens}}) > 0.85$) and low disagreement ($D(\mathbf{x}) < 0.15$) are classified immediately without further processing. In Stage~2 (ADS calculation), remaining samples trigger multi-signal suspicion scoring; if the ADS score falls below the detection threshold, the sample is classified as clean. In Stage~3 (deep verification), highly suspicious samples undergo comprehensive analysis including two-stage adaptive detection, attack-adaptive weighting, and confidence calibration before final classification. This staged approach is intended to avoid expensive $O(d^2)$ Mahalanobis computation for the majority of traffic; empirical routing distributions are reported in Section~\ref{sec:results}.


\subsection*{Attack-Driven Suspicion Score (ADS)}
\label{sec:method_ads}

The ADS calculator combines three complementary signals entropy, disagreement, and anomaly through learned weights to detect adversarial manipulation.

\subsubsection{Multi-Signal Detection}

We compute three detection signals intended to capture distinct aspects of adversarial behavior.

\textit{Entropy} measures prediction uncertainty from ensemble probabilities using Shannon entropy~\cite{shannon1948mathematical}:
\begin{equation}
H(\mathbf{x}) = -\sum_{c=0}^{C-1} p_{\text{ens}}^{(c)} \log p_{\text{ens}}^{(c)}
\label{eq:entropy}
\end{equation}

\textit{Disagreement} captures cross-model prediction variance as defined in Equation~\ref{eq:disagreement}.

\textit{Anomaly score} measures feature-space deviation using Mahalanobis distance~\cite{mahalanobis1936generalized} from the benign training distribution $\mathcal{N}(\boldsymbol{\mu}, \boldsymbol{\Sigma})$:
\begin{equation}
A(\mathbf{x}) = \sqrt{(\mathbf{x} - \boldsymbol{\mu})^T \boldsymbol{\Sigma}^{-1} (\mathbf{x} - \boldsymbol{\mu})}
\label{eq:anomaly}
\end{equation}

\subsubsection{Weighted Combination and Learning}

The ADS combines normalized signals $\hat{H}, \hat{D}, \hat{A} \in [0,1]$ through learned weights:
\begin{equation}
\text{ADS}(\mathbf{x}; \mathbf{w}) = \alpha \cdot \hat{H}(\mathbf{x}) + \beta \cdot \hat{D}(\mathbf{x}) + \gamma \cdot \hat{A}(\mathbf{x})
\label{eq:ads}
\end{equation}

where $\mathbf{w} = [\alpha, \beta, \gamma]$ with $\alpha + \beta + \gamma = 1$. We learn three sets of weights. Generic weights $\mathbf{w}_{\text{gen}}$ are optimized for combined adversarial detection across all attack types:
\begin{equation}
\mathbf{w}_{\text{gen}} = \arg\max_{\mathbf{w}} \text{AUC}(\text{ADS}(\mathcal{D}_{\text{clean}}; \mathbf{w}), \text{ADS}(\mathcal{D}_{\text{adv}}; \mathbf{w}))
\label{eq:generic_weights}
\end{equation}

Attack-specific weights $\mathbf{w}_k$ are optimized per attack type $k$ using the same objective restricted to $\mathcal{D}_k$. Category-level weights $\mathbf{w}_{\text{gradient}}$ and $\mathbf{w}_{\text{dist}}$ are then derived by averaging attack-specific weights within each attack category (gradient-based and distribution shift, respectively).

\subsubsection{Two-Stage Adaptive Detection}

Two-stage detection combines generic inference with category-specific refinement (Algorithm~\ref{alg:two_stage}). In Stage~1, the system computes a generic suspicion score using $\mathbf{w}_{\text{gen}}$ and simultaneously infers the attack category from the normalized anomaly score: samples with $\hat{A}(\mathbf{x}) > \tau_{\text{anomaly}} = 0.50$ are categorized as distribution shift attacks, while remaining samples are categorized as gradient-based. In Stage~2, the system recomputes the suspicion score using category-specific weights $\mathbf{w}_{\hat{c}}$, producing a refined detection decision. The rationale is that category-level refinement may compensate for weak individual attack-specific weights by leveraging shared structure within attack families (evaluated in Section~\ref{sec:results}).

\begin{algorithm}[!t]
\caption{Two-Stage Adaptive Detection}
\label{alg:two_stage}
\begin{algorithmic}[1]
\REQUIRE Input sample $\mathbf{x}$, generic weights $\mathbf{w}_{\text{gen}}$, category weights $\{\mathbf{w}_{\text{gradient}}, \mathbf{w}_{\text{dist}}\}$, threshold $\tau_{\text{anomaly}}=0.50$
\ENSURE Binary detection $y_{\text{detect}}$, inferred category $\hat{c}$

\STATE \textbf{// Stage 1: Generic Inference and Category Classification}
\STATE Compute signals: $H(\mathbf{x}), D(\mathbf{x}), A(\mathbf{x})$
\STATE Normalize: $\hat{H}, \hat{D}, \hat{A} \leftarrow$ min-max normalize to $[0,1]$
\STATE $\text{ADS}_{\text{gen}} \leftarrow \mathbf{w}_{\text{gen}}^T \cdot [\hat{H}, \hat{D}, \hat{A}]$

\STATE \textbf{// Infer attack category from anomaly score}
\IF{$\hat{A}(\mathbf{x}) > \tau_{\text{anomaly}}$}
    \STATE $\hat{c} \leftarrow$ distribution
\ELSE
    \STATE $\hat{c} \leftarrow$ gradient
\ENDIF

\STATE \textbf{// Stage 2: Specialized Refinement}
\STATE Select weights: $\mathbf{w}_{\text{spec}} \leftarrow \mathbf{w}_{\hat{c}}$
\STATE $\text{ADS}_{\text{refined}} \leftarrow \mathbf{w}_{\text{spec}}^T \cdot [\hat{H}, \hat{D}, \hat{A}]$

\STATE \textbf{// Final detection}
\IF{$\text{ADS}_{\text{refined}} > \tau_{\text{detect}}$}
    \STATE $y_{\text{detect}} \leftarrow 1$ \textit{(adversarial)}
\ELSE
    \STATE $y_{\text{detect}} \leftarrow 0$ \textit{(clean)}
\ENDIF

\RETURN $y_{\text{detect}}, \hat{c}$
\end{algorithmic}
\end{algorithm}


\subsection*{Attack-Adaptive Model Weighting}
\label{sec:method_adaptive}

Attack-adaptive weighting adjusts ensemble model weights based on the inferred attack category, aiming to leverage model-specific strengths against different threat types.

\subsubsection{Weight Learning}

For each attack category $c \in \{\text{gradient}, \text{distribution}\}$, we learn model weights $\mathbf{v}_c = [v_1^{(c)}, v_2^{(c)}, \ldots, v_6^{(c)}]$ through performance-based optimization:
\begin{equation}
v_i^{(c)} = \frac{(\text{Acc}_i^{(c)})^3}{\sum_{j=1}^{6} (\text{Acc}_j^{(c)})^3}
\label{eq:model_weights}
\end{equation}

where $\text{Acc}_i^{(c)}$ is model $i$'s accuracy on attack category $c$. The cubic exponent is intended to create sharp weight distributions that favor high-performing models for each category.

\subsubsection{Confidence-Based Blending}

To mitigate potential over-reliance on the inferred category when confidence is low, we blend adaptive weights with uniform weights $\mathbf{v}_{\text{uniform}} = [1/6, \ldots, 1/6]$. Confidence is derived from the margin between the anomaly score and the category threshold: $\text{conf}(\hat{c}) = \min(1, |\hat{A}(\mathbf{x}) - \tau_{\text{anomaly}}| / 0.12)$. When $\text{conf}(\hat{c}) < 0.75$, we apply $\mathbf{v}_{\text{final}} = 0.5 \cdot \mathbf{v}_{\hat{c}} + 0.5 \cdot \mathbf{v}_{\text{uniform}}$; otherwise, $\mathbf{v}_{\text{final}} = \mathbf{v}_{\hat{c}}$.

The adaptive ensemble prediction is then:
\begin{equation}
\hat{y}_{\text{adaptive}} = \arg\max_{c} \sum_{i=1}^{6} v_i^{(\hat{c})} \cdot p_i^{(c)}
\label{eq:adaptive_prediction}
\end{equation}


\subsection*{Dual-Output Architecture}
\label{sec:method_dual_output}

AMDS produces two complementary outputs: adversarial detection ($y_{\text{detect}} \in \{0,1\}$) indicating whether the input appears to have been manipulated, with confidence $\text{confidence}_{\text{detect}} = \min(1, \text{ADS}_{\text{refined}} / \tau_{\text{detect}})$; and attack classification ($y_{\text{class}} \in \{0,\ldots,C-1\}$) identifying the traffic type, with confidence $\text{confidence}_{\text{class}} = \max(\mathbf{p}_{\text{ens}})$.


\subsection*{Training and Inference Pipelines}
\label{sec:method_integration}

\subsubsection{Training Pipeline}

Algorithm~\ref{alg:training} describes the AMDS training procedure, which proceeds in four stages: (1)~training six ensemble models independently on clean data (target: ${>}$95\% accuracy); (2)~generating adversarial samples across gradient-based (FGSM, PGD, C\&W, SPSA) and distribution shift (Injection, Morphing) attacks, validated at ${>}$50\% ASR; (3)~learning detection weights through AUC optimization at generic, attack-specific, and category levels; and (4)~tuning detection and routing thresholds on validation data ($\tau_{\text{detect}}$ calibrated for 10\% FPR).

\begin{algorithm}[!t]
\caption{AMDS Training Pipeline}
\label{alg:training}
\begin{algorithmic}[1]
\REQUIRE Clean training data $\mathcal{D}_{\text{train}}$, validation data $\mathcal{D}_{\text{val}}$
\ENSURE Trained models $\mathcal{M}$, learned weights $\{\mathbf{w}_{\text{gen}}, \mathbf{w}_k, \mathbf{v}_c\}$, thresholds $\{\tau_*\}$

\STATE \textbf{// Stage 1: Baseline Model Training}
\FOR{each model type $i \in \{1,\ldots,6\}$}
    \STATE Train $M_i$ on $\mathcal{D}_{\text{train}}$ for classification accuracy
    \STATE Validate $M_i$ on $\mathcal{D}_{\text{val}}$ (target: $>$95\% accuracy)
\ENDFOR

\STATE \textbf{// Stage 2: Adversarial Sample Generation}
\STATE Generate gradient attacks: $\mathcal{D}_{\text{FGSM}}, \mathcal{D}_{\text{PGD}}, \mathcal{D}_{\text{CW}}, \mathcal{D}_{\text{SPSA}}$
\STATE Generate distribution attacks: $\mathcal{D}_{\text{Injection}}, \mathcal{D}_{\text{Morphing}}$
\STATE Combine: $\mathcal{D}_{\text{adv}} = \bigcup_k \mathcal{D}_k$
\STATE Validate attack effectiveness (target: ASR $>$50\%)

\STATE \textbf{// Stage 3: Weight Learning}
\STATE Learn generic weights: $\mathbf{w}_{\text{gen}} \leftarrow \arg\max_{\mathbf{w}} \text{AUC}(\mathcal{D}_{\text{val}}, \mathcal{D}_{\text{adv}}; \mathbf{w})$
\FOR{each attack type $k$}
    \STATE Learn attack-specific: $\mathbf{w}_k \leftarrow \arg\max_{\mathbf{w}} \text{AUC}(\mathcal{D}_{\text{val}}, \mathcal{D}_k; \mathbf{w})$
\ENDFOR
\STATE Average within categories: $\mathbf{w}_{\text{gradient}}, \mathbf{w}_{\text{dist}}$
\FOR{each category $c \in \{\text{gradient}, \text{distribution}\}$}
    \STATE Learn model weights: $v_i^{(c)} \leftarrow (\text{Acc}_i^{(c)})^3 / \sum_j (\text{Acc}_j^{(c)})^3$
\ENDFOR

\STATE \textbf{// Stage 4: Threshold Tuning}
\STATE Tune $\tau_{\text{detect}}$ for 10\% FPR on $\mathcal{D}_{\text{val}}$
\STATE Tune $\tau_{\text{conf}}, \tau_{\text{disagree}}$ for cascade routing
\STATE Tune $\tau_{\text{anomaly}}$ for optimal category inference accuracy

\RETURN $\mathcal{M}, \{\mathbf{w}_{\text{gen}}, \mathbf{w}_k, \mathbf{v}_c\}, \{\tau_*\}$
\end{algorithmic}
\end{algorithm}

\subsubsection{Inference Pipeline}

Algorithm~\ref{alg:inference} describes the complete inference procedure integrating all four components. The worst-case computational complexity is $O(6 \cdot C_{\text{model}} + d^2)$, where $C_{\text{model}}$ is individual model inference cost and $d^2$ arises from the Mahalanobis distance computation. The cascade router is designed to reduce average-case complexity by resolving the majority of samples at Stage~1 without incurring the $O(d^2)$ cost.

\begin{algorithm}[!t]
\caption{AMDS Inference Pipeline}
\label{alg:inference}
\begin{algorithmic}[1]
\REQUIRE Input sample $\mathbf{x}$, trained models $\mathcal{M}$, weights $\{\mathbf{w}_*, \mathbf{v}_*\}$, thresholds $\{\tau_*\}$
\ENSURE Adversarial detection $y_{\text{detect}}$, attack classification $y_{\text{class}}$, confidences

\STATE \textbf{// Stage 1: Ensemble Prediction}
\FOR{each model $M_i \in \mathcal{M}$}
    \STATE $\hat{y}_i, \mathbf{p}_i \leftarrow M_i(\mathbf{x})$
\ENDFOR
\STATE $\hat{y}_{\text{ens}} \leftarrow$ majority vote from $\{\hat{y}_1, \ldots, \hat{y}_6\}$
\STATE $D(\mathbf{x}) \leftarrow$ compute disagreement (Equation~\ref{eq:disagreement})

\STATE \textbf{// Cascade Routing: Stage 1 Check}
\IF{$\max(\mathbf{p}_{\text{ens}}) > \tau_{\text{conf}}$ \textbf{and} $D(\mathbf{x}) < \tau_{\text{disagree}}$}
    \STATE \textbf{return} $y_{\text{detect}}=0$, $y_{\text{class}}=\hat{y}_{\text{ens}}$ \textit{// Fast path}
\ENDIF

\STATE \textbf{// Stage 2: ADS Calculation \& Two-Stage Detection}
\STATE $y_{\text{detect}}, \hat{c} \leftarrow$ TwoStageDetection($\mathbf{x}$) \textit{// Algorithm~\ref{alg:two_stage}}

\STATE \textbf{// Cascade Routing: Stage 2 Check}
\IF{$y_{\text{detect}} = 0$}
    \STATE \textbf{return} $y_{\text{detect}}=0$, $y_{\text{class}}=\hat{y}_{\text{ens}}$
\ENDIF

\STATE \textbf{// Stage 3: Deep Verification with Attack-Adaptive Weighting}
\STATE Compute confidence: $\text{conf}(\hat{c}) \leftarrow \min(1, |\hat{A}(\mathbf{x}) - \tau_{\text{anomaly}}| / 0.12)$
\IF{$\text{conf}(\hat{c}) > 0.75$}
    \STATE $\mathbf{v}_{\text{final}} \leftarrow \mathbf{v}_{\hat{c}}$
\ELSE
    \STATE $\mathbf{v}_{\text{final}} \leftarrow 0.5 \cdot \mathbf{v}_{\hat{c}} + 0.5 \cdot \mathbf{v}_{\text{uniform}}$
\ENDIF
\STATE $\hat{y}_{\text{adaptive}} \leftarrow \arg\max_c \sum_{i=1}^{6} v_i^{(\hat{c})} \cdot p_i^{(c)}$

\STATE \textbf{// Dual Output}
\STATE $y_{\text{class}} \leftarrow \hat{y}_{\text{adaptive}}$
\STATE $\text{confidence}_{\text{detect}} \leftarrow \min(1, \text{ADS}_{\text{refined}} / \tau_{\text{detect}})$
\STATE $\text{confidence}_{\text{class}} \leftarrow \max(\mathbf{p}_{\text{ens}})$

\RETURN $y_{\text{detect}}, y_{\text{class}}, \text{confidence}_{\text{detect}}, \text{confidence}_{\text{class}}$
\end{algorithmic}
\end{algorithm}


\section{Experimental Evaluation}
\label{sec:experiments}

This section describes the datasets, attack generation, baselines, and evaluation protocol used to assess AMDS.


\subsection*{Datasets}

We evaluate AMDS on two benchmark datasets. \textbf{CSE-CIC-IDS2018}~\cite{sharafaldin2018cicids2017} serves as the primary evaluation dataset, containing realistic network traffic across 7 classes (Benign, Botnet, Brute-force, DDoS, DoS, Infiltration, Web attack) with 77 flow-level features. We partition the data into 50,000 training, 5,000 validation, and a held-out test set of 5,000 benign plus 1,000 adversarial samples per attack type (500 for C\&W due to computational cost), using stratified sampling throughout. \textbf{UNSW-NB15}~\cite{moustafa2015unsw} provides cross-dataset evaluation with 190 features and 9 attack categories mapped to 5 classes aligned with CSE-CIC-IDS2018. The substantially higher dimensionality (2.47$\times$) enables investigation of how attack effectiveness may scale with feature space. Identical experimental protocols are applied across both datasets.


\subsection*{Data Preprocessing}

Standard preprocessing is applied: duplicate removal, missing value imputation, and zero-mean unit-variance scaling fitted on training data only to prevent leakage. Benign distribution parameters $\boldsymbol{\mu}$ and $\boldsymbol{\Sigma}$ are computed from clean training traffic for the Mahalanobis-based anomaly score (Equation~\ref{eq:anomaly}).


\subsection*{Attack Generation}

We generate seven adversarial attacks spanning two categories using the Adversarial Robustness Toolbox (ART) v1.13~\cite{nicolae2018adversarial} for gradient-based attacks and NumPy for distribution shift attacks.

\textit{Gradient-based attacks} (five types): FGSM~\cite{goodfellow2014explaining} ($\epsilon{=}0.02$, single-step $L_\infty$); PGD-$L_\infty$~\cite{madry2017towards} ($\epsilon{=}0.02$, 20 steps, $\alpha{=}0.005$); PGD-$L_2$~\cite{madry2017towards} ($\epsilon{=}0.02$, 20 steps, $L_2$-constrained); C\&W-$L_2$~\cite{carlini2017towards} (20 binary search iterations, $\kappa{=}0$, 500 samples); and SPSA (100 queries, $\delta{=}0.01$, black-box gradient estimation).

\textit{Distribution shift attacks} (two types): Injection adds per-feature Gaussian noise $\mathcal{N}(0, 0.05 \cdot \sigma_i)$, while Morphing applies systematic offset $x_i' = x_i + 0.05 \cdot \sigma_i$, where $\sigma_i$ is the training standard deviation.

All attacks are validated for effectiveness (target: ASR${>}$50\%) before inclusion. PGD-$L_2$ achieves a lower ASR of 16.6\% on CSE-CIC-IDS2018 due to the relatively small $\epsilon$ in 77-dimensional space; we retain it to examine detector behavior under weak-attack conditions.


\subsection*{Baselines}

The ensemble comprises six classifiers: Decision Tree, Random Forest~\cite{breiman2001random}, XGBoost~\cite{chen2016xgboost}, LightGBM, Logistic Regression, and a Multi-Layer Perceptron, spanning tree-based, boosting, linear, and neural paradigms. All models use standard hyperparameters validated for ${>}$95\% clean accuracy; detailed configurations are provided in supplementary material.\footnote{Decision Tree (depth 10), Random Forest (100 trees), XGBoost (100 estimators, lr 0.1), LightGBM (100 leaves, lr 0.1), Logistic Regression ($C{=}1.0$, L2), MLP (layers: 100, 50; ReLU).}

We compare AMDS against four baselines using identical training data and evaluation protocols: (1)~\textit{Standard Ensemble} with uniform voting; (2)~\textit{Adversarial Training (AT) Ensemble} retrained with PGD adversarial training~\cite{madry2017towards} (50\% clean, 50\% adversarial; $\epsilon{=}0.02$, 10 steps), averaged over three random seeds with 95\% bootstrap confidence intervals; (3)~\textit{Single Best Model} (XGBoost); and (4)~\textit{Uniform ADS}, i.e., AMDS with generic weights only.


\subsection*{Evaluation Metrics}

We report AUC as the primary detection metric (threshold-independent), along with accuracy, F1-score, and per-attack breakdowns at the operating threshold (calibrated for 10\% FPR). System efficiency is measured via per-sample latency and cascade routing distribution. All metrics include 95\% bootstrap confidence intervals (1,000 iterations, percentile method).


\subsection*{Experimental Protocol}

Evaluation proceeds in three phases: (1)~\textit{component validation}, assessing individual detection signals, cascade routing, two-stage detection, and attack-adaptive weighting independently; (2)~\textit{full system evaluation} on held-out data (2,000 clean + 500 adversarial per attack type = 5,500 samples) against all baselines, including cross-dataset evaluation on UNSW-NB15; and (3)~\textit{ablation study} removing individual components to quantify their contributions.\footnote{Implementation: Python 3.9, scikit-learn 1.2, XGBoost 1.7, PyTorch 2.0; Intel Xeon E5-2680 v4, 128\,GB RAM, NVIDIA V100 GPU. Weight optimization via SciPy SLSQP. All seeds fixed (42) for reproducibility. Code and models to be released upon acceptance.}


\section{Results}
\label{sec:results}

We report results in three parts: primary evaluation on CSE-CIC-IDS2018, cross-dataset generalization on UNSW-NB15, and robustness analysis including adaptive adversary evaluation.

\subsection{Primary Evaluation on CSE-CIC-IDS2018}
\label{sec:results_primary}

\subsubsection{Attack-Specific Weight Patterns}
\label{sec:results_weights}

Table~\ref{tab:attack_specific_weights} presents learned attack-specific weights for entropy ($\alpha$), disagreement ($\beta$), and anomaly ($\gamma$) across six attack types. PGD-L$_2$ is excluded due to weak attack effectiveness (ASR = 16.6\%), which yields near-random detection weights (AUC = 0.502); its two-stage results appear in Table~\ref{tab:two_stage_comparison}.

\begin{table*}[!t]
\centering
\caption{Attack-Specific Learned Weights and Detection Performance}
\label{tab:attack_specific_weights}
\begin{tabular}{lccccccc}
\hline
\textbf{Attack} & \textbf{Category} & \boldmath{$\alpha$} & \boldmath{$\beta$} & \boldmath{$\gamma$} & \textbf{AUC} & \textbf{Dominant} \\
 & & \textbf{(Entropy)} & \textbf{(Disagree)} & \textbf{(Anomaly)} & & \textbf{Signal} \\
\hline
FGSM & Gradient & 0.260 & \textbf{0.528} & 0.212 & 0.743 & Disagreement \\
PGD-L$_\infty$ & Gradient & 0.269 & \textbf{0.465} & 0.267 & 0.759 & Disagreement \\
C\&W & Gradient & 0.306 & \textbf{0.434} & 0.261 & 0.748 & Disagreement \\
SPSA & Gradient & 0.293 & \textbf{0.448} & 0.259 & 0.758 & Disagreement \\
\hline
Injection & Distribution & \textbf{0.433} & 0.360 & 0.207 & 0.896 & Entropy \\
Morphing & Distribution & 0.454 & 0.013 & \textbf{0.533} & 0.988 & Anomaly \\
\hline
\textbf{Gradient avg.} & & 0.282 & \textbf{0.469} & 0.250 & 0.752 & \textbf{Uncertainty} \\
\textbf{Distribution avg.} & & 0.444 & 0.187 & 0.370 & 0.942 & \textbf{Heterogeneous} \\
\hline
\end{tabular}

\par\noindent{\footnotesize
Gradient-based attacks exhibit disagreement-dominant patterns ($\beta \approx 0.47$). Distribution attacks show heterogeneous patterns: Morphing is anomaly-dominant ($\gamma=0.533$, AUC = 0.988), while Injection is entropy-dominant ($\alpha=0.433$, AUC = 0.896).
PGD-L$_2$ excluded due to weak attack effectiveness (16.6\% ASR); its near-random weights (AUC = 0.502) likely reflect insufficient perturbation strength rather than a detection limitation.}
\end{table*}

Gradient-based attacks consistently exhibit disagreement-dominant patterns ($\beta$ ranging from 0.434 to 0.528), while distribution shift attacks show heterogeneous signatures: Injection appears entropy-dominant ($\alpha = 0.433$) and Morphing anomaly-dominant ($\gamma = 0.533$). These patterns suggest that different attack categories may require specialized detection strategies. Generic weights optimized across all attacks yield $\alpha = 0.163$, $\beta = 0.333$, $\gamma = 0.504$ with 97.1\% AUC. Category-level weights produce: gradient ($\alpha = 0.294$, $\beta = 0.452$, $\gamma = 0.253$, AUC = 0.702) and distribution ($\alpha = 0.443$, $\beta = 0.187$, $\gamma = 0.370$, AUC = 0.942).

\subsubsection{Two-Stage Detection Efficacy}
\label{sec:results_twostage}

Table~\ref{tab:two_stage_comparison} compares three detection strategies across all seven attack types.

\begin{table*}[!t]
\centering
\caption{Two-Stage Detection Comparison (AUC)}
\label{tab:two_stage_comparison}
\begin{tabular}{lcccccc}
\hline
\textbf{Attack} & \textbf{Generic-Only} & \textbf{Attack-Specific} & \textbf{Two-Stage} & \boldmath{$\Delta$} \textbf{vs Generic} & \boldmath{$\Delta$} \textbf{vs Specific} \\
\hline
FGSM & 0.545 & 0.743 & \textbf{0.943} & +0.398 & +0.200 \\
PGD-L$_\infty$ & 0.552 & 0.759 & \textbf{0.943} & +0.391 & +0.184 \\
PGD-L$_2$ & 0.218 & 0.502 & \textbf{0.885} & +0.666 & +0.383 \\
C\&W & 0.565 & 0.748 & \textbf{0.945} & +0.379 & +0.197 \\
SPSA & 0.551 & 0.758 & \textbf{0.943} & +0.392 & +0.185 \\
Injection & 0.652 & 0.896 & \textbf{0.966} & +0.313 & +0.070 \\
Morphing & 0.967 & 0.988 & \textbf{0.971} & +0.004 & $-0.017$ \\
\hline
\textbf{Average (7 attacks)} & \textbf{0.579} & \textbf{0.771} & \textbf{0.942} & \textbf{+0.363} & \textbf{+0.172} \\
\hline
\end{tabular}
\vspace{4pt}

\par\noindent{\footnotesize
\textit{Generic-Only}: generic weights ($\alpha = 0.163$, $\beta = 0.333$, $\gamma = 0.504$).
\textit{Attack-Specific-Only}: individual weights per attack (seven weight sets).
\textit{Two-Stage}: generic screening + category-level refinement (two weight sets).
PGD-L$_2$ benefits most from two-stage detection (+0.383 vs specific-only), suggesting that category-level refinement may compensate for weak individual attack weights.}
\end{table*}

Two-stage detection achieves 0.942 average AUC, outperforming generic-only (0.579, +0.363) and attack-specific-only (0.771, +0.172). PGD-L$_2$ shows the largest improvement (+0.383 vs specific-only), rising from near-random individual detection (AUC = 0.502) to strong category-level detection (AUC = 0.885), suggesting that category-level refinement can compensate for weak individual attack weights.

Category inference achieves 100\% accuracy across all seven attacks. Distribution attacks produce mean anomaly scores of 0.68--0.85 versus 0.15--0.25 for gradient attacks, enabling threshold-based inference ($\tau_{\mathrm{anomaly}} = 0.50$). While this clean separability holds on the evaluated attacks, it should not be assumed to generalize to all possible attack types without further validation.

\subsubsection{Cascade Computational Characteristics}
\label{sec:results_cascade}

Table~\ref{tab:cascade_efficiency} compares full AMDS processing against cascade routing.

\begin{table}[!t]
\centering
\caption{Cascade Router Performance and Efficiency}
\label{tab:cascade_efficiency}
\begin{tabular}{lccc}
\hline
\textbf{Metric} & \textbf{Baseline} & \textbf{Cascade} & \textbf{Ratio} \\
 & \textbf{(Full AMDS)} & \textbf{(Adaptive)} & \\
\hline
Throughput (samples/sec) & 6{,}172 & 4{,}989 & 0.81$\times$ \\
Latency (ms/sample) & 0.16 & 0.20 & +0.04 \\
Overall Accuracy & 84.9\% & 84.9\% & 0.0\,pp \\
Clean Accuracy & 96.3\% & 96.3\% & 0.0\,pp \\
\hline
\multicolumn{4}{l}{\textbf{Routing Distribution:}} \\
Stage 1 (fast screening) & 100\% & 76.5\% & $-23.5$\% \\
Stage 2 (ADS calculation) & 100\% & 23.4\% & $-76.6$\% \\
Stage 3 (deep verification) & 100\% & 9.0\% & $-91.0$\% \\
\hline
\multicolumn{4}{l}{\textbf{Computational Operations:}} \\
Mahalanobis calculations & 100\% & 23.5\% & $-76.5$\% \\
Attack-adaptive weighting & 100\% & 9.0\% & $-91$\% \\
\hline
\end{tabular}
\vspace{4pt}

\par\noindent{\footnotesize
Cascade achieves 76.5\% Stage~1 routing with zero accuracy degradation.
Throughput overhead (0.81$\times$) results from routing logic; Mahalanobis calculations decrease by 76.5\% and adaptive weighting by 91\%.}
\end{table}

Cascade routing resolves 76.5\% of samples at Stage~1, with 23.4\% proceeding to Stage~2 and 9.0\% to Stage~3 (deep verification), maintaining identical accuracy (84.9\% overall, 96.3\% clean) in both configurations. The 0.81$\times$ throughput ratio reflects routing logic overhead, though Mahalanobis calculations decrease by 76.5\% and attack-adaptive weighting by 91\%. The cascade benefit thus lies in computational redistribution rather than raw throughput: deployments where expensive matrix operations dominate cost may benefit from this trade-off, while throughput-critical environments may prefer the non-cascaded configuration.

\subsubsection{Baseline Comparisons and Attack-Adaptive Weighting}
\label{sec:results_baseline}

\begin{figure*}[t]
\centering
\includegraphics[width=0.95\textwidth]{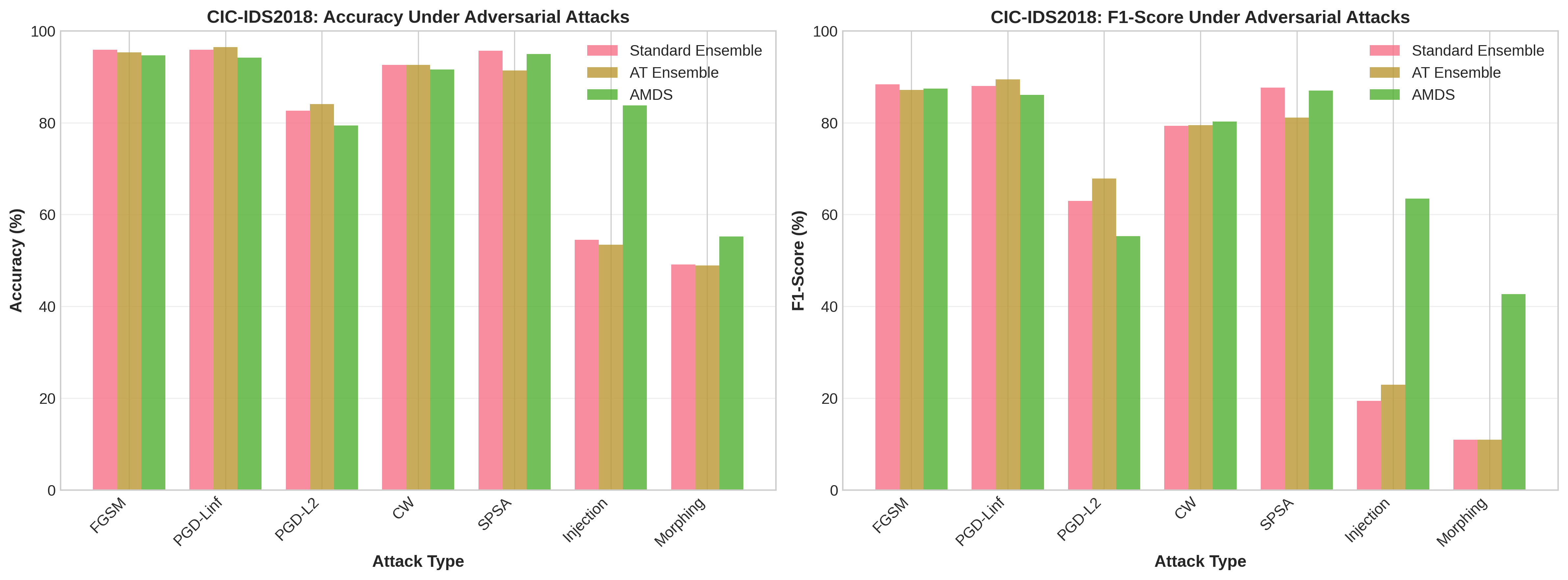}
\caption{Per-attack performance comparison on CSE-CIC-IDS2018. AMDS achieves substantial improvements on distribution shift attacks (+29.3\,pp Injection, +6.1\,pp Morphing vs Standard), with small regression on gradient attacks ($-1.6$\,pp average). Adversarial training appears to provide minimal benefit over standard training.}
\label{fig:cic_comparison}
\end{figure*}

Table~\ref{tab:baseline_comparison} presents the primary baseline comparison. Figure~\ref{fig:cic_comparison} shows the per-attack breakdown.

\begin{table*}[!t]
\centering
\caption{Comprehensive Baseline Comparison on CSE-CIC-IDS2018}
\label{tab:baseline_comparison}
\begin{tabular}{lccccccccc}
\hline
\textbf{Attack} & \textbf{Cat.} & \textbf{Standard} & \textbf{AT} & \textbf{AMDS} & \boldmath{$\Delta$} \textbf{vs Std} & \boldmath{$\Delta$} \textbf{vs AT} & \textbf{Std F1} & \textbf{AT F1} & \textbf{AMDS F1} \\
\hline
\multicolumn{10}{l}{\textit{Gradient-Based Attacks}} \\
FGSM & G & 95.9\% & 95.3\% & 94.7\% & $-1.2$\,pp & $-0.6$\,pp & 88.3\% & 87.1\% & 87.4\% \\
PGD-L$_\infty$ & G & 95.9\% & 96.5\% & 94.2\% & $-1.7$\,pp & $-2.3$\,pp & 88.0\% & 89.4\% & 86.1\% \\
PGD-L$_2$ & G & 82.6\% & 84.1\% & 79.4\% & $-3.2$\,pp & $-4.7$\,pp & 63.0\% & 67.8\% & 55.3\% \\
C\&W & G & 92.6\% & 92.6\% & 91.6\% & $-1.0$\,pp & $-1.0$\,pp & 79.3\% & 79.5\% & 80.3\% \\
SPSA & G & 95.7\% & 91.4\% & 95.0\% & $-0.7$\,pp & $+3.6$\,pp & 87.6\% & 81.1\% & 87.0\% \\
\hline
\textbf{Gradient avg.} & & 92.5\% & 92.0\% & 91.0\% & $-1.6$\,pp & $-1.0$\,pp & 81.2\% & 81.0\% & 79.2\% \\
\hline
\multicolumn{10}{l}{\textit{Distribution Shift Attacks}} \\
Injection & D & 54.5\% & 53.4\% & \textbf{83.8\%} & \textbf{+29.3\,pp} & \textbf{+30.4\,pp} & 19.4\% & 22.9\% & \textbf{63.5\%} \\
Morphing & D & 49.1\% & 48.9\% & \textbf{55.2\%} & \textbf{+6.1\,pp} & \textbf{+6.3\,pp} & 11.0\% & 10.9\% & \textbf{42.7\%} \\
\hline
\textbf{Distribution avg.} & & 51.8\% & 51.2\% & \textbf{69.5\%} & \textbf{+17.7\,pp} & \textbf{+18.4\,pp} & 15.2\% & 16.9\% & \textbf{53.1\%} \\
\hline
\textbf{Overall} & & 80.9\% & 80.3\% & \textbf{84.8\%} & \textbf{+3.9\,pp} & \textbf{+4.5\,pp} & 62.4\% & 62.7\% & \textbf{71.7\%} \\
 & & & & \footnotesize{(74.1--92.5\%)} & & & & & \footnotesize{(60.9--83.1\%)} \\
\textbf{Clean} & & 98.3\% & 99.3\% & 96.3\% & $-2.0$\,pp & $-3.0$\,pp & 49.6\% & 33.2\% & 32.7\% \\
 & & & & \footnotesize{(95.3--97.0\%)} & & & & & \\
\hline
\end{tabular}
\vspace{4pt}

\par\noindent{\footnotesize
95\% bootstrap CIs ($n = 1{,}000$) in parentheses. Wide overall CIs (74.1--92.5\%) reflect heterogeneity across attack types. AT trained with PGD ($\epsilon = 0.02$, three seeds). PGD-L$_2$ regression reflects weak ASR (16.6\%), included for completeness.}
\end{table*}

AMDS achieves 84.8\% overall accuracy (+3.9\,pp vs Standard, +4.5\,pp vs AT) and 71.7\% F1-score (+9.0\,pp vs AT). Adversarial training appears to provide minimal benefit over standard training ($-0.6$\,pp accuracy, +0.3\,pp F1). On gradient attacks, AMDS shows small regression ($-1.0$\,pp vs AT average), which may be expected given that baselines already handle these attacks well (92.0--92.5\% average). On distribution attacks, AMDS achieves substantial gains (+18.4\,pp vs AT), with Injection showing +30.4\,pp accuracy and +40.6\,pp F1 improvement. These results suggest that adversarial training may be insufficient against distribution shift threats, while attack-specific optimization can provide meaningful benefit.

Category inference achieves perfect precision, recall, and F1 (1.000) for both gradient and distribution categories, enabling reliable selection of category-specific model weights. Clean accuracy shows a modest regression (96.3\% vs 98.3\% Standard, 99.3\% AT), representing a trade-off that appears favorable given the +9.0\,pp overall F1 gain.

\subsubsection{Component Ablation Analysis}
\label{sec:results_ablation}

Table~\ref{tab:ablation_study} presents ablation results.

\begin{table}[!t]
\centering
\caption{Ablation Study: Component Contribution Analysis}
\label{tab:ablation_study}
\begin{tabular}{lccc}
\hline
\textbf{Configuration} & \textbf{Overall} & \textbf{Overall} & \textbf{Clean} \\
 & \textbf{Accuracy} & \textbf{F1} & \textbf{Accuracy} \\
\hline
\textbf{Full AMDS} & \textbf{84.9\%} & \textbf{73.0\%} & \textbf{96.3\%} \\
\hline
\multicolumn{4}{l}{\textit{Ablated Configurations:}} \\
Generic weights only & 81.4\% & 63.2\% & 98.1\% \\
\quad Component contribution & $-3.5$\,pp & $-9.8$\,pp & $+1.8$\,pp \\
Uniform model voting & 81.1\% & 61.0\% & 99.5\% \\
\quad Component contribution & $-3.8$\,pp & $-12.0$\,pp & $+3.2$\,pp \\
No attack-specific weights & 81.2\% & 62.1\% & 98.1\% \\
\quad Component contribution & $-3.7$\,pp & $-10.9$\,pp & $+1.8$\,pp \\
\hline
\end{tabular}
\vspace{4pt}

\par\noindent{\footnotesize
All components contribute $\geq$3\,pp accuracy and $\geq$9.8\,pp F1. Ablated configurations show increased clean accuracy (98.1--99.5\%), indicating a trade-off between clean performance and adversarial robustness. Test set: 2{,}000 clean + 500 per attack = 5{,}500 samples.}
\end{table}

All three components contribute meaningfully: two-stage detection +3.5\,pp, attack-adaptive weighting +3.8\,pp, and attack-specific weights +3.7\,pp. F1-score contributions are larger (+9.8\,pp to +12.0\,pp), suggesting a stronger impact on precision--recall balance than on raw accuracy. Ablated configurations show increased clean accuracy (98.1--99.5\% vs 96.3\%), revealing an explicit trade-off: AMDS sacrifices 1.8--3.2\,pp of clean accuracy to gain 3.5--3.8\,pp of overall accuracy and 9.8--12.0\,pp of F1-score under adversarial conditions.

\subsection{Cross-Dataset Generalization (UNSW-NB15)}
\label{sec:results_crossdataset}

Table~\ref{tab:cross_dataset} presents cross-dataset evaluation results.

\begin{table*}[!t]
\centering
\caption{Cross-Dataset Generalization: CSE-CIC-IDS2018 vs.\ UNSW-NB15}
\label{tab:cross_dataset}
\begin{tabular}{lcccc}
\hline
\textbf{Metric} & \textbf{CSE-CIC-IDS2018} & \textbf{UNSW-NB15} & \boldmath{$\Delta$} & \textbf{Explanation} \\
\hline
\multicolumn{5}{l}{\textit{Dataset Characteristics}} \\
Features & 77 & 190 & +113 & 2.47$\times$ more dimensions \\
Attack Budget ($\epsilon$) & 0.02 & 0.02 & 0.0 & Same nominal budget \\
Effective Attack Strength & 1.0$\times$ & ${\sim}1.57\times$ & +0.57$\times$ & Scales with $\sqrt{d}$ \\
\hline
\multicolumn{5}{l}{\textit{Baseline Performance}} \\
Baseline Accuracy & 80.7\% & 52.9\% & $-27.8$\,pp & Attack strength exceeds baseline capacity \\
Clean Accuracy & 98.1\% & 98.5\% & +0.4\,pp & Models function on clean data \\
\hline
\multicolumn{5}{l}{\textit{AMDS Performance}} \\
AMDS Accuracy & 84.9\% & 52.7\% & $-32.2$\,pp & Limited by baseline degradation \\
AMDS F1-Macro & 73.0\% & 37.6\% & $-35.4$\,pp & Baseline floor effect \\
$\Delta$ Accuracy vs Baseline & +4.2\,pp & $-0.2$\,pp & $-4.4$\,pp & Baseline likely too weak \\
$\Delta$ F1 vs Baseline & +11.0\,pp & +0.7\,pp & $-10.3$\,pp & Minimal improvement \\
\hline
\multicolumn{5}{l}{\textit{Validation: Weaker Attacks ($\epsilon = 0.01$)}} \\
$\Delta$ Accuracy ($\epsilon = 0.01$) & -- & +0.5\,pp & -- & Improvement partially recovers \\
$\Delta$ F1 ($\epsilon = 0.01$) & -- & +1.9\,pp & -- & Stronger F1 gain \\
\hline
\end{tabular}
\vspace{4pt}

\par\noindent{\footnotesize
Effective attack strength appears to scale as $\sqrt{d}$: $\sqrt{190}/\sqrt{77} \approx 1.57\times$, so $\epsilon = 0.02$ on UNSW may be effectively equivalent to $\epsilon \approx 0.031$ on CIC. With weaker attacks ($\epsilon = 0.01$), AMDS recovers positive improvement (+0.5\,pp accuracy, +1.9\,pp F1), though baseline accuracy remains low (51.5\%).}
\end{table*}

UNSW-NB15 baseline accuracy (52.9\%) is substantially lower than CSE-CIC-IDS2018 (80.7\%), while clean accuracy remains high on both datasets ($>$96\%), indicating that baseline degradation stems from attack strength rather than model inadequacy. AMDS achieves $-0.2$\,pp on UNSW versus +4.2\,pp on CIC, suggesting that when attacks overwhelm baseline models, adaptive defenses may have limited room for improvement. This performance gap appears attributable to dimensionality-dependent attack scaling: under identical $\epsilon = 0.02$, UNSW-NB15's 190 features likely experience $\sim$1.57$\times$ stronger effective perturbation than CIC's 77 features ($\sqrt{d}$ scaling). Validation with weaker attacks ($\epsilon = 0.01$) partially recovers positive improvement (+0.5\,pp accuracy, +1.9\,pp F1), consistent with this interpretation. These results point to a practical deployment consideration: AMDS effectiveness appears to depend on baseline model competence, and practitioners deploying in high-dimensional feature spaces ($>$150 features) should consider scaling perturbation budgets proportionally or applying dimensionality reduction.

\subsection{Robustness Analysis}
\label{sec:results_robustness}

\subsubsection{Adaptive Adversary Evaluation}

Table~\ref{tab:adaptive_adversary} presents results for adaptive attacks with white-box knowledge of the AMDS architecture.

\begin{table}[!t]
\centering
\caption{Robustness Against Adaptive Adversaries}
\label{tab:adaptive_adversary}
\begin{tabular}{lccc}
\hline
\textbf{Attack Type} & \textbf{Standard} & \textbf{AMDS} & \textbf{Detection} \\
 & \textbf{Ensemble} & \textbf{Accuracy} & \textbf{Rate (ADS)} \\
\hline
\multicolumn{4}{l}{\textit{Standard Attacks (Non-Adaptive)}} \\
FGSM & 95.1\% & 94.6\% & 2.4\% \\
PGD-L$_\infty$ & 95.5\% & 93.4\% & 1.8\% \\
PGD-L$_2$ & 83.4\% & 80.5\% & 14.0\% \\
\hline
\multicolumn{4}{l}{\textit{Adaptive Attacks (ADS-Aware)}} \\
Baseline Adaptive & 97.9\% & 96.1\% & 5.0\% \\
Improved Adaptive & 95.8\% & \textbf{94.4\%} & 3.1\% \\
\hline
\textbf{Performance Drop} & -- & \textbf{$-1.7$\,pp} & -- \\
\textbf{Attack Success Rate} & -- & \textbf{4.2\%} & -- \\
\hline
\end{tabular}
\vspace{4pt}

\par\noindent{\footnotesize
Adaptive attacks minimize ADS score while maintaining misclassification under full white-box knowledge. \textit{Baseline Adaptive}: benign start, random direction search. \textit{Improved Adaptive}: FGSM start, gradient approximation. ADS reduction is 0.0\% for both variants.}
\end{table}

We evaluate two adaptive variants: Baseline Adaptive (starting from benign samples with random direction search) and Improved Adaptive (starting from FGSM-perturbed samples with gradient approximation). AMDS maintains 94.4\% accuracy under the improved adaptive attack ($-1.7$\,pp vs baseline adaptive), with 4.2\% attack success rate. Neither variant achieves any ADS score reduction (0.0\%), suggesting that simultaneously minimizing entropy, disagreement, and anomaly while maintaining misclassification may impose conflicting optimization objectives. The improved variant achieves higher success (4.2\% vs 2.1\%) by leveraging FGSM-perturbed starting points, though this rate remains well below typical ASR for standard attacks ($>$50\%). We note that this evaluation covers two adaptive variants and does not constitute a formal robustness guarantee; a sufficiently sophisticated adversary might identify strategies not explored here.

\subsubsection{Statistical Significance}

All metrics include 95\% bootstrap confidence intervals (1{,}000 iterations, percentile method). Overall accuracy is 85.01\% [74.09\%, 92.46\%], F1-macro is 73.08\% [60.86\%, 83.06\%], and clean accuracy is 96.25\% [95.32\%, 97.00\%]. The wide overall CI reflects heterogeneity across attack types; paired bootstrap tests indicate significance at $p < 0.05$ for improvements over both baselines.


\section{Conclusion}
\label{sec:conclusion}

This paper presented AMDS, a multi-stage adversarial defense system for network intrusion detection that learns attack-specific detection strategies rather than applying uniform defenses. Evaluation on CSE-CIC-IDS2018, cross-dataset validation on UNSW-NB15, and robustness testing against adaptive adversaries suggest both practical effectiveness and identifiable deployment boundaries.

AMDS outperforms adversarially trained ensembles by +4.5~pp accuracy and +9.0~pp F1-score, while adversarial training itself appears to provide negligible benefit over standard ensembles ($-$0.6~pp accuracy). The learned weight patterns indicate that gradient attacks tend to exhibit disagreement-dominant signatures ($\beta \approx 0.47$), whereas distribution attacks show heterogeneous patterns---anomaly-dominant for Morphing, entropy-dominant for Injection. Two-stage adaptive detection achieves 94.2\% AUC across seven attack types by operating at an intermediate category-level granularity, which appears to avoid both the oversmoothness of generic detection and the overfitting tendencies of individual attack-specific detection. Under white-box adaptive attacks that explicitly minimize detection scores, AMDS maintains 94.4\% accuracy with a 4.2\% attack success rate, though this evaluation does not constitute a formal robustness guarantee.

Cross-dataset evaluation on UNSW-NB15 reveals that AMDS effectiveness likely depends on baseline model competence. Attack strength appears to scale with feature dimensionality ($\propto \sqrt{d}$), causing UNSW-NB15's 190-dimensional feature space to experience effectively $\sim$1.57$\times$ stronger perturbations than CSE-CIC-IDS2018's 77-dimensional space under identical budgets. When this scaling drives baselines toward chance-level performance, adaptive defenses including but not limited to AMDS may have limited room for improvement.

Several limitations should be noted. First, AMDS requires adversarial examples during weight learning, assuming partial knowledge of likely attack types; how gracefully performance degrades for entirely unseen attack families remains an open question. Second, the binary category inference (gradient vs.\ distribution) may not capture the full spectrum of real-world adversarial diversity, and the 100\% inference accuracy observed here should not be extrapolated to finer-grained taxonomies where inter-class boundaries may be less distinct. Third, learned weights remain static during deployment; although adaptive adversary evaluation suggests robustness to explicit ADS minimization, a sufficiently persistent adversary could potentially craft specialized evasion strategies over time. Fourth, the wide bootstrap confidence intervals (overall accuracy 95\% CI: 74.1--92.5\%) reflect substantial performance heterogeneity across attack types, and per-attack breakdowns (Table~IV) should be consulted alongside summary statistics for a complete picture of system capabilities.

These limitations suggest several directions for future investigation. Online weight adaptation could address the static-weight vulnerability, though it introduces the challenge of distinguishing genuine attack evolution from adversarial manipulation of the learning signal. Extending the attack taxonomy beyond binary categorization may enable finer-grained optimization, subject to empirical validation of inference accuracy at higher granularity. Transfer to other adversarial ML domains malware detection, fraud detection, autonomous perception would test whether the category-level separability observed in NIDS generalizes more broadly. Finally, integration with certified robustness techniques could complement AMDS's empirical effectiveness with formal guarantees in settings where provable bounds are required.

Overall, these findings suggest that attack-specific optimization may represent a more effective path to adversarial robustness in network intrusion detection than adversarial training alone, achieving 85.0\% overall accuracy (95\% CI: 74.1--92.5\%) with empirically grounded deployment considerations.

\bibliographystyle{IEEEtran}
\bibliography{references}

\end{document}